\documentclass[aps,prb,reprint,superscriptaddress]{revtex4-1}
\usepackage{bm}
\usepackage{graphicx}
\usepackage{graphics}
\usepackage{amsmath}
\usepackage{amsfonts}
\usepackage{amssymb}
\usepackage{epstopdf}
\usepackage{hyperref}
\usepackage{soul}
\usepackage{tikz}
\usepackage{hyperref}
\hypersetup{colorlinks=true,citecolor=blue,linkcolor=blue,filecolor=blue,urlcolor=blue}
\begin{document}
\title{Tunable spin-polarized edge transport in inverted quantum-well junctions}
\author{Dimy Nanclares}
\affiliation{Instituto de F\'{\i}sica, Universidade de S\~{a}o Paulo,
C.P.\ 66318, 05315--970 S\~{a}o Paulo, SP, Brazil}
\author{Leandro R.\ F.\ Lima}
\affiliation{Instituto de F\'{\i}sica, Universidade Federal Fluminense, 
24210-346 Niter\'oi, Brazil}
\author{Caio H. Lewenkopf}
\affiliation{Instituto de F\'{\i}sica, Universidade Federal Fluminense, 
24210-346 Niter\'oi, Brazil}
\author{Luis G.\ G.\ V.\ Dias da Silva}
\affiliation{Instituto de F\'{\i}sica, Universidade de S\~{a}o Paulo,
C.P.\ 66318, 05315--970 S\~{a}o Paulo, SP, Brazil}
\date{\today}
\begin{abstract}
Inverted HgTe/CdTe quantum wells have been used as a platform for the realization of 2D 
topological insulators, bulk insulator materials with spin-helical metallic edges states protected by time-reversal symmetry. This work investigates the spectrum and the charge transport in HgTe/CdTe quantum well 
\textit{junctions} both in the topological regime and in the absence of time-reversal symmetry. 
We model the system using the BHZ effective Hamiltonian and compute the transport properties 
using recursive Green's functions with a 
finite differences' method.
Specifically, we have studied the material's spatially-resolved conductance in a set-up with a 
gated central region, forming monopolar (n-n$^{\prime}$-n) and heteropolar (n-p-n, n-TI-n) double 
junctions, which have been recently realized in experiments. We find regimes in which the edge states 
carry spin-polarized currents in the central region even in the presence of a small magnetic field, 
which breaks TRS. More interestingly, the conductance displays spin-dependent, Fabry-Per\'ot-like 
oscillations as a function of the central gate voltage producing tunable, fully spin-polarized currents 
through the device.

\end{abstract}
%
\maketitle
%
\section{Introduction}\label{sec:Intro}

The role of topology in the properties of electronic systems has gained renewed attention over the last decade with the discovery of several materials that support topologically-protected surface and edge states, dubbed ``topological insulators" (TIs). \cite{Hasan:3045:2010,Qi:1057:2011,Ando:J.Phys.Soc.Jpn.:102001:2013}  Particular attention has been given to 2D topological insulators, where the quantum spin Hall (QSH) effect    \cite{Kane:Phys.Rev.Lett.:146802:2005} allows for edge electron transport through spin-polarized helical edge states. The theoretical proposal\cite{Bernevig:1757:2006} and later observation\cite{Konig:766:2007} of the QSH effect in HgTe/CdTe quantum wells has triggered intense activity in the study of these systems.

The electronic current in 2D topological insulators is carried by edge states protected by time-reversal symmetry (TRS). In the absence of TRS, backscattering between the edge modes becomes allowed and the gaplessness of the edge states is no longer guaranteed. It has been argued\cite{Konig:766:2007} that even a small magnetic field is sufficient to open a gap in the edge states, thereby suppressing  edge transport in 2D topological insulators. The argument supporting this view comes from the early experiments\cite{Konig:766:2007} on HgTe/CdTe quantum wells  which show that the magnetoconductance shows a cusp-like feature at zero field, quickly decaying as the field increases. Such behavior, however, can only be accounted for when a rather strong disorder (of the order of the bulk gap) is considered \cite{Maciejko:Phys.Rev.B:155310:2010} in addition to TRS breaking. 

In many situations, however, breaking TRS does not imply a suppression of edge transport channels 
in these systems. Several theoretical studies \cite{Schmidt:Phys.Rev.B:241306:2009, Tkachov:Phys.Rev.Lett.:104:166803:2010, Buttner:NatPhys:418--422:2011, Chen:Phys.Rev.B:125401:2012, Scharf:Phys.Rev.B:75418:2012, Raichev:Phys.Rev.B:85:045310:2012, Durnev:Phys.Rev.B:93:075434:2016} 
as well as experimental evidence \cite{Gusev:121302:2011, Gusev:Phys.Rev.B:81311:2013, 
Gusev:Phys.Rev.Lett.:76805:2013} point to a scenario where edge transport in HgTe/CdTe {quantum wells (QWs)} is quite 
relevant up to magnetic fields of a few Tesla. 
For instance, theory predicts \cite{Tkachov:Phys.Rev.Lett.:104:166803:2010, 
Chen:Phys.Rev.B:125401:2012, Scharf:Phys.Rev.B:75418:2012} a transition from helical {QSH} 
to chiral QHE edge states at a critical field of a few Tesla.

A recent theoretical study on HgTe/CdTe QWs has shown that for small system sizes 
the edge states remain unaffected by a combination of moderate disorder and weak 
magnetic fields.\cite{Essert:2DMaterials:024005:2015}
The transport properties change in long samples when considering charge 
puddles.  \cite{Essert:2DMaterials:024005:2015} 
This kind of disorder and the corresponding local potential fluctuations have been extensively 
studied in graphene systems. \cite{Martin2008,Mucciolo2010,DasSarma2011} It has been found that charge puddles give rise to a disordered
landscape of $p$-$n$ junctions that are key to understand the low energy electronic transport
in realistic graphene samples.\cite{Cheianov2007,Lima2016,Fan2017} 

Some recent theoretical \cite{Zhang:NewJournalofPhysics:083058:2010} and experimental\cite{Gusev:Phys.Rev.Lett.:76805:2013,Piatrusha:arXiv:1703.09816::2017} works  have
probed the transport properties of 
heteropolar lateral junctions in HgTe/CdTe QWs in the inverted regime. 
Reference~\onlinecite{Gusev:Phys.Rev.Lett.:76805:2013} in particular investigates electronic transport of 
\textit{double} junction systems by applying a gate voltage $V_g$ in the central region of a HgTe quantum-well 
Hall bar. By varying $V_g$, the system can be tuned from an 
n-n$^{\prime}$-n type junction (Fermi energy lying in the electron-like states of the junction) to 
n-p-n (Fermi energy lying in the hole-like states of the junction). When $V_g$ is tuned close to 
the charge {neutrality} point, $V_g\!=\!V^{\rm CN}_g$, the Fermi energy lies near the gap 
of the central region and the transport across 
the junction is expected
to be dominated, in the absence of magnetic field, by QSH edge states (``n-TI-n'').

The results presented in Ref.~\onlinecite{Gusev:Phys.Rev.Lett.:76805:2013} show that, in the presence of a 
strong perpendicular magnetic field ($B\agt 7$ T) the system enters the quantum Hall regime. 
The longitudinal conductance displays plateaus consistent with those expected for graphene junctions in the QHE regime:\cite{Abanin:Science:641:2007,Williams:Science:638--641:2007} $2 e^2/h$ in a monopolar n-n$^\prime$-n junction ($V_g \! \gg \! V^{\rm CN}_g$) and $e^2/2h$ in the bipolar
n-p-n junction ($V_g \! \ll \! V^{\rm CN}_g$). In the n-TI-n configuration ($V_g \approx V^{\rm CN}_g$), a non-quantized
conductance value  was measured. More intriguingly, well-defined conductance plateaus have not been observed
for weaker magnetic fields. \cite{Gusev:Phys.Rev.Lett.:76805:2013}
This regime has no clear interpretation yet, calling for further theoretical investigation.

In this paper, we present a simplified model to describe the weak field  limit (non-QHE regime) of HgTe/CdTe QW junctions.
We compute the transport properties of pristine HgTe/CdTe QWs  and homopolar and heteropolar double junctions by combining a 
discrete model Hamiltonian with the recursive Green's functions method (RGF). \cite{Lewenkopf2013}
For concreteness, we consider the Bernevig-Hughes-Zhang (BHZ) model \cite{Bernevig:1757:2006} in the presence of a 
perpendicular magnetic field. \cite{Scharf:Phys.Rev.B:75418:2012, Scharf:Phys.Rev.B:235433:2015} 

We calculate the space-resolved transmission across different types of junctions 
(n-n$^{\prime}$-n, n-edge-n, n-p-n) 
as a function of a gate voltage $V_g$ applied at the system central region
and of an external magnetic field. The latter is believed to destroy the topological
protection, since it allows for backscattering in the spin-polarized edge states. 
Interestingly, our results show that, for fields up to a few Tesla, there is always a range
of $V_g$ where edge transmission in the central region dominates the transport properties. 
We refer to this configuration as 
an ``n-edge-n" junction. 

We show that some of the transport features of the studied $n$-TI (or $p$-TI) junctions bear similarities with the case of graphene junctions where the transmission to a region where the transport is forbidden can be understood in terms of ``snake-like" states at the interface.\cite{Carmier2010,Carmier2011}

One of our main results is that, in an n-edge-n junction, the combined effect of quantum interference from reflection at the junction barriers and edge-state backscattering due to the breaking of TRS creates a spin-dependent Fabry-P\'erot pattern in the transmission amplitudes. These gate-controlled oscillations are strong enough to provide fully spin-polarized currents across the junction. 

The paper is organized as follows. In Sec.\ \ref{sec:modelmethods} we present the BHZ model 
used to describe the HgTe junctions and discuss the recursive Green's functions approach 
employed to investigate the local currents in the system. Our numerical results are presented in 
Sec.\ \ref{sec:Results}, where we 
study
the effect of 
a perpendicular external magnetic field on
the transport properties across the junction. Finally, we present our concluding remarks in Sec.\ \ref{sec:Conclusions}.

\section{Model and Methods}
\label{sec:modelmethods}

We describe the physical properties of HgTe/CdTe QWs at low energies and zero 
magnetic field using the 4-band BHZ Hamiltonian \cite{Bernevig:1757:2006} 
\begin{align}
    \label{eq:Hamiltonian}
    \hat{H}= &\mathcal{C} \textbf{1} 
        + \mathcal{M} \Gamma_5 
    -\frac{\left(\mathcal{D}\textbf{1} 
        + \mathcal{B}\Gamma_5\right)}{\hbar^2} \left( \hat{p}^2_x + \hat{p}^2_y \right) \nonumber\\ 
    &-\frac{\mathcal{A}\Gamma_1}{\hbar} \hat{p}_x 
    + \frac{\mathcal{A}\Gamma_2}{\hbar} \hat{p}_y,
\end{align}
where $\Gamma_1$, $\Gamma_2$ and $\Gamma_5$ are $4 \times 4$ matrices spanning the basis 
$\left\{ 
\left|E\uparrow\right\rangle, 
\left|H\uparrow\right\rangle, 
\left|E\downarrow\right\rangle, 
\left|H\downarrow\right\rangle 
\right\}$, 
that can be expressed in terms of Pauli matrices $\sigma_j$, namely
\begin{align}
    \Gamma_1 &= \left(\begin{array}{cc}
                                    \sigma_x & 0 \\
                                    0 & -\sigma_x
                                    \end{array}\right), 
    \Gamma_2 = \left(\begin{array}{cc}
                                    -\sigma_y & 0 \\
                                    0 & -\sigma_y
                                    \end{array}\right), 
                                    \nonumber\\
    \Gamma_5 &= \left(\begin{array}{cc}
                                    \sigma_z & 0 \\
                                    0 & \sigma_z
                                    \end{array}\right),
\end{align}
and $\mathbf 1$ is the identity.
The numerical parameters $\mathcal{A},\mathcal{B},\mathcal{C},\mathcal{D}$ depend on system properties such 
as the QW thickness.

We caution
that the indices ``$\uparrow$" and ``$\downarrow$" indicate degenerate Kramers pairs related by 
TRS in the low-energy effective model obtained from the original $\bf{k} \cdot \bf{p}$ 6-band model 
for HgTe near the $\Gamma$ ($\bf{k}=0)$ point.\cite{Bernevig:1757:2006} 
In this sense, the latter are not ``pure" spin 1/2 states since 
the $H$-states carry contributions from p-type heavy-hole bands with spin $J_z\!=\!\pm 3/2$. 
However, to a good approximation,  these states represent spin 1/2 states related by TRS 
\cite{Buttner:NatPhys:418--422:2011,Scharf:Phys.Rev.B:75418:2012} and we will treat them as 
such in the present work.
In addition, we will neglect inversion-breaking terms \cite{Durnev:Phys.Rev.B:93:075434:2016} 
which give rise to coupling between the spin up and the spin down sectors.

\begin{figure}[t]
    \includegraphics[width=0.45\columnwidth]{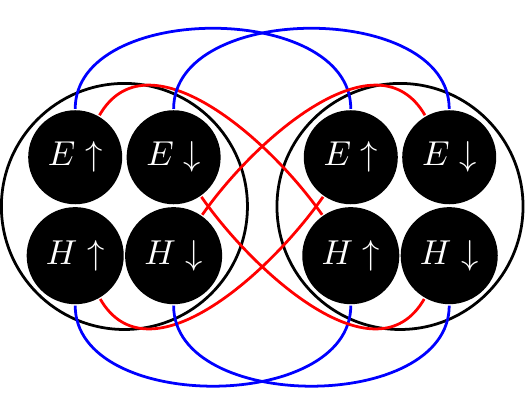}
    \includegraphics[width=0.45\columnwidth]{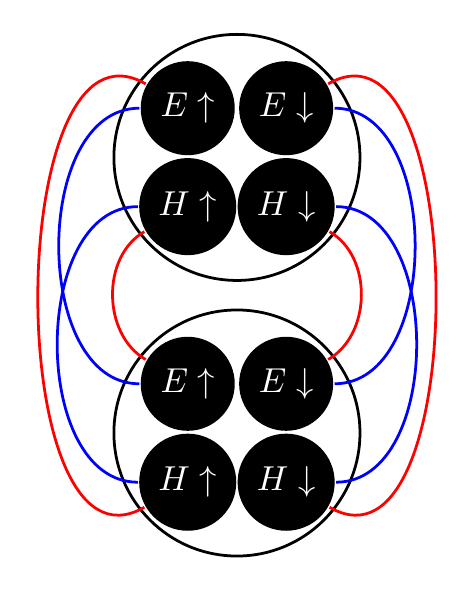}
    \caption{Sketch of the hopping 
    matrix element structure in our discretized model. Each site (large circle) has a 
                  fourfold orbital structure with states (black circles) of the basis $\left\{ 
\left|E\uparrow\right\rangle, 
\left|H\uparrow\right\rangle, 
\left|E\downarrow\right\rangle, 
\left|H\downarrow\right\rangle 
\right\}$.
The lines represent the non-vanishing hoppings
between states that belong to different sites.
No hopping between states in the same site is allowed.  
 Left: hoppings between orbitals belonging to the sites at ($n,m$) and ($n+1,m$).
    Right: hoppings between ($n,m$) and ($n,m+1$). }
    \label{fig:orbitals}
\end{figure}

The numerical calculation of the QW transport properties follows the prescription of 
Ref.~\onlinecite{Scharf:Phys.Rev.B:75418:2012}.
We discretize the 4-component spinor $\Psi(x,y)$ in a square lattice of spacing $a$ in both $x$ 
and $y$ directions.
The spinor $\Psi(x,y)$ becomes $\Psi_{n,m}$ where $x=na$ and $y=ma$ and $n$ and $m$ are integer.
Figure~\ref{fig:orbitals} shows the orbital structure of the hopping matrix elements.
We note that the hopping terms between electron and hole states are nonzero only if 
the spin projection is preserved.

\begin{figure}[t]
    \centering
    \includegraphics[width=1.00\columnwidth]{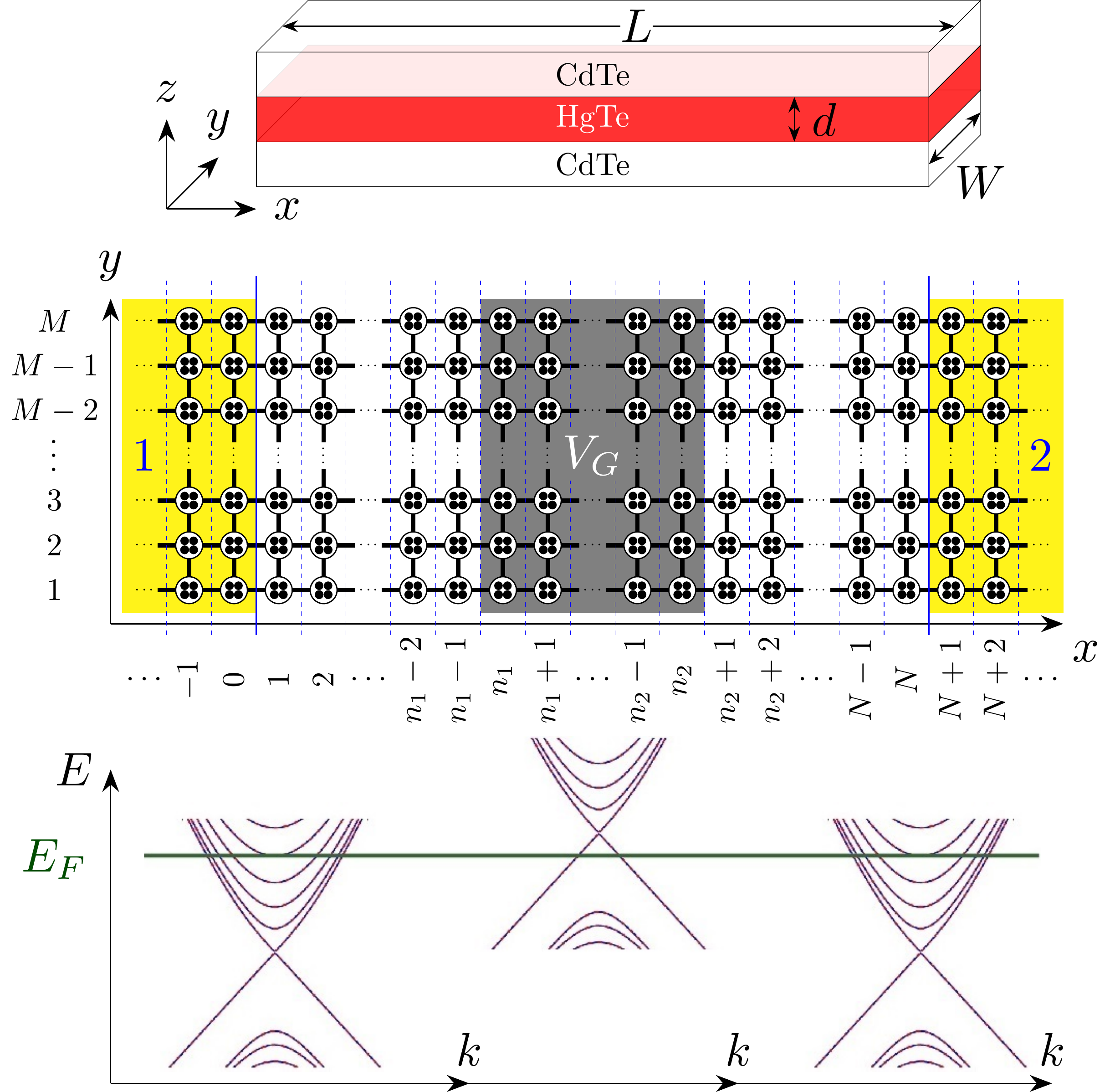}
\caption{
Top: Sketch of the HgTe/CdTe quantum well of thickness $d$, width $W$, and length $L$. 
Middle: Sketch of the real space discretization. 
$N$ is the number of vertical slices in the central region and $M$ is the number of sites in each slice. 
Each site contains 4 orbitals represented by black dots. 
The yellow shaded regions correspond to the ``left" and ``right" leads. 
The gray area represents the HgTe regions over which a gate voltage $V_g$ is applied.
Bottom: Local band structure shift due to the effect of a negative gate voltage $V_g$. This configuration 
corresponds to an n-edge-n (or n-TI-n at $B\!=\!0$) junction.
}
\label{fig:tbstructure}
\end{figure}

Using two and three points derivatives for the momenta discretization
\begin{align}
    \label{eq:DiscBeq0}
    p_x  \Psi(x,y)  &
    \rightarrow -\frac{i\hbar}{2a}\left(\Psi_{n+1}^{m}-\Psi_{n-1}^{m}\right), \\ \nonumber
    p^2_x\Psi(x,y) &
    \rightarrow -\frac{\hbar^2}{a^2}\left(\Psi_{n+1}^{m}-2\Psi_{n}^{m}+ \Psi_{n-1}^{m}\right), \\ \nonumber
    p_y  \Psi(x,y)  &
    \rightarrow -\frac{i\hbar}{2a}\left(\Psi_{n}^{m+1}-\Psi_{n}^{m-1}\right), \\ \nonumber
    p^2_y\Psi(x,y) &
    \rightarrow -\frac{\hbar^2}{a^2}\left(\Psi_{n}^{m+1}-2\Psi_{n}^{m}+ \Psi_{n}^{m-1}\right),
\end{align}
the eigenvalue problem $\hat H\Psi(x,y)=E\Psi(x,y)$, becomes
\begin{align}
    E \Psi_n^m =
    H_{n,n  }^{m,m  }&\Psi_{n  }^{m  } +
    H_{n,n-1}^{m,m  } \Psi_{n-1}^{m  } +
    H_{n,n+1}^{m,m  } \Psi_{n+1}^{m  }  \nonumber \\ & +
    H_{n,n  }^{m,m-1} \Psi_{n}^{m-1} +
    H_{n,n  }^{m,m+1} \Psi_{n}^{m+1}
    \label{eigenproblem}
\end{align}
where
\begin{align}
    H_{n,n  }^{m,m  } &= 
    \bigg[ 
        \mathcal{C} \textbf{1} + \mathcal{M} \Gamma_5 
        - 4\frac{\left(\mathcal{D} \textbf{1} + \mathcal{B}\Gamma_5\right)}{a^2}   
                + \frac{\mu_B B\Gamma^z_g}{2}
    \bigg],
        \label{hlocal}
     \\ 
    H_{n,n+1}^{m,m  } &=
     \left[
        \frac{\left(\mathcal{D} \textbf{1} + \mathcal{B}\Gamma_5\right)}{a^2} 
                + \frac{i \mathcal{A} \Gamma_1}{2a}  
    \right] e^{ima^2(eB/\hbar)},
        \label{hx} %
     \\ 
    H_{n,n}^{m,m+1} &=
    \left[
    \frac{\left(\mathcal{D} \textbf{1} +\mathcal{B}\Gamma_5\right)}{a^2} 
    - \frac{i \mathcal{A} \Gamma_2}{2a}  
    \right],
    \label{hy}
    \\
    H_{n,n-1}^{m,m} &= \left(H_{n,n+1}^{m,m}\right)^\dagger,
    \ \ \ \
    H_{n,n}^{m,m-1} = \left(H_{n,n}^{m,m+1}\right)^\dagger.
    \label{hydagger}
\end{align}
The above model Hamiltonian accounts for the presence of an external magnetic field 
perpendicular to the QW ($\mathbf B=B \mathbf{\hat z}$) by means of the 
gauge ${\bf A}({\bf r})=-By\hat{{\bf x}}$ and by a Zeeman term \cite{Buttner:NatPhys:418--422:2011} 
$\mu_B B\Gamma^z_g/2$ 
where $\Gamma^z_g =\mbox{diag}(g_e,g_h,g_e,g_h)$ contains the effective $g$-factors for electrons $g_e$ and holes $g_h$
and $\mu_B$ is the Bohr magneton.
The Peierls phase $(e/\hbar)\int_{(n,m)}^{(n+1,m)} \mathbf A \cdot d\mathbf l = ma^2(eB/\hbar)$ 
modifies the hopping matrix elements between the sites $(n,m)$ and $(n+1,m)$ in Eq.~(\ref{hx}).

Comparisons with full 8-band $\bf{k} \cdot \bf{p}$ 
calculations  show that this low-energy model offers a good description for HgTe/CdTe QWs
near the $\Gamma$ point for magnetic fields up 
to $B \sim 2$ T.\cite{Schmidt:Phys.Rev.B:241306:2009}

\begin{figure}[t]
\centering
\includegraphics[width=1.0\columnwidth]{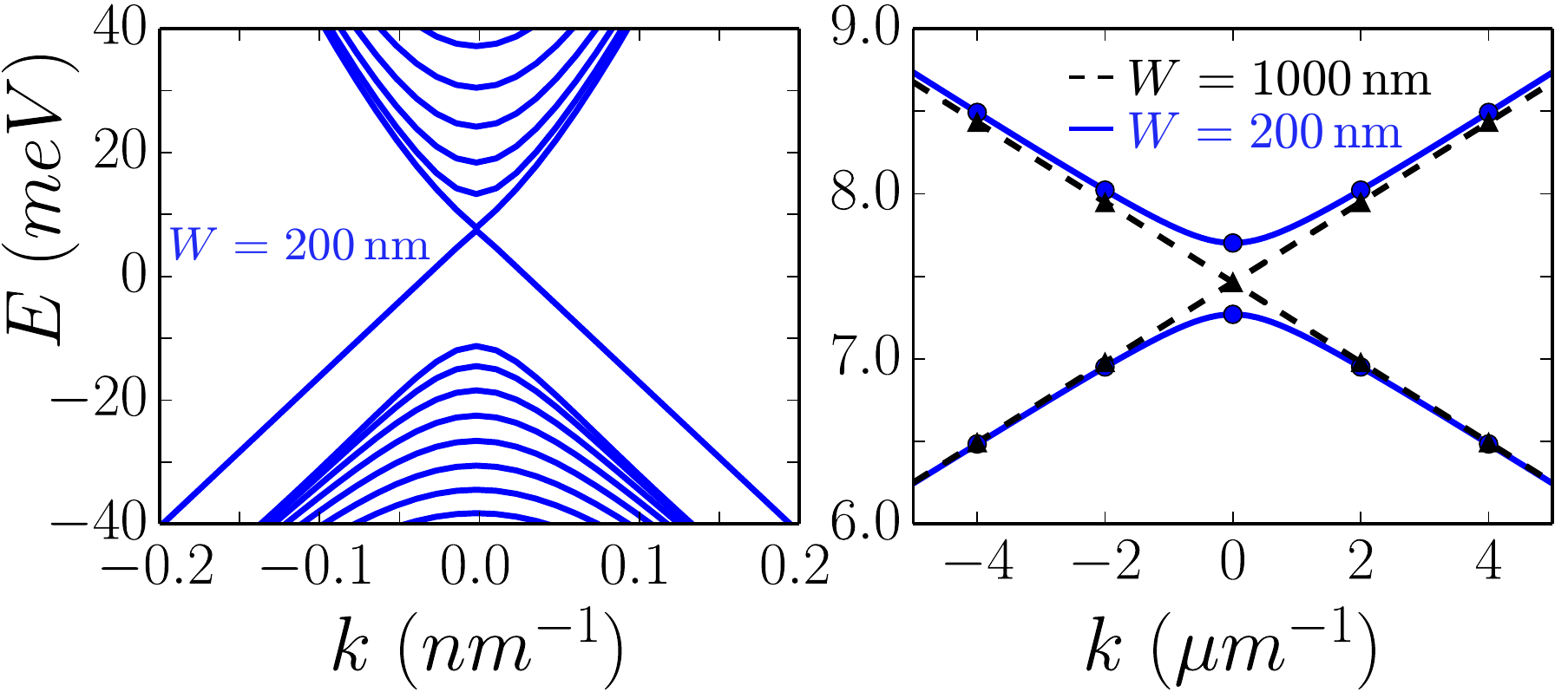}
\caption{ (Color online) 
(a) Spectrum of an 
infinite HgTe/CdTe QW of width $W=200$ nm. (b) Detail of the edge states near $k\!=\!0$ showing a 
small gap for $W=200$ nm  (blue squares) and essentially no gap for $W=1000$ nm (black triangles).
Symbols correspond to our finite differences calculation and the solid lines to analytic results of Ref.\ 
\onlinecite{Zhou:Phys.Rev.Lett.:246807}.
}
\label{fig:dispersion1D}
\end{figure}

We address the transport properties of a QW of thickness $d$, width $W$, and length $L$. 
The system is attached to left and right semi-infinite leads, aligned to its longitudinal direction, 
parallel to the $x$-axis, as we illustrate in Fig.~\ref{fig:tbstructure}. 
For computational convenience the QW region is divided into transverse slices that 
are labeled by $n$ ranging from $n_1$ to $n_2$, see Fig.~\ref{fig:tbstructure}. 
We model homopolar and heteropolar junctions by introducing a gate voltage $V_g$ acting on 
the system central region, corresponding to a term 
\begin{align}
\label{eq:Vg}
    \left[H_G\right]_{n,n}^{m,m} = eV_g \mathbf 1
\end{align}
for $n_1 \le n \le n_2$. In this way, we approximate the model Hamiltonian of 
Eq.~\eqref{eq:Hamiltonian} by a discrete Hamiltonian on a square lattice of 
dimensions $M\times N$, where $W=Ma$ and $L=Na$, containing $4MN$ 
orbitals in the central region. 

The choice of the lattice parameter $a$ is a compromise between computational cost and accuracy.
The value of $a$ is fixed as follows: We solve the eigenproblem in Eq.~(\ref{eigenproblem}) using 
periodic boundary conditions in the $x$ direction for a chosen width $W$ in the $y$ direction.
The reduced eigenvalue problem reads
\begin{align}
    E \mathbf \Psi_n =
    \mathbf H_{n,n  }\mathbf \Psi_{n  } +
    \mathbf H_{n,n-1}\mathbf \Psi_{n-1} +
    \mathbf H_{n,n+1}\mathbf \Psi_{n+1},
    \label{eigenproblemribbon}
\end{align}
where $\mathbf H_{n,n'}$ has the block structure 
$\left[\mathbf H_{n,n}\right]^{m,m'} = H_{n,n}^{m,m'}\left(\delta_{m',m}+\delta_{m',m-1}+\delta_{m',m+1}\right)$ and $\left[\mathbf H_{n,n\pm 1}\right]^{m,m'} = H_{n,n\pm 1}^{m,m}\delta_{m',m}$. 
Here $H_{n,n'}^{m,m'}$ is a $4 \times 4$ matrix, given by Eqs.~\eqref{hlocal} to \eqref{hydagger}. 

Due to translational invariance $\mathbf \Psi_n$ can be written as $\mathbf \Psi_n= 
\boldsymbol \psi e^{ik_xna}$, where $\boldsymbol \psi$ has $4M$ components. 
For a given width $W$ we choose $a$ by requiring an accuracy of $10^{-2}$ meV in the energy 
gap as compared to the analytical results obtained in Ref.~\onlinecite{Zhou:Phys.Rev.Lett.:246807}.
In practice, we fix $M=200$ for all calculations, which sets a lattice parameter $a$ for a given 
width $W$. We find that this procedure satisfies the required accuracy for systems with $W \alt 2 \mu$m.

Figure~\ref{fig:dispersion1D} shows valence and conduction bands of two infinite HgTe/CdTe QWs of 
widths $W=200$ nm and $W=1000$ nm obtained using the material parameters of a $d\!=\!7$ nm
 thick QW \cite{Zhou:Phys.Rev.Lett.:246807,Konig:J.Phys.Soc.Jpn.:031007:2008}
$\mathcal{A}=364.5\ \text{meV~nm}$, 
$\mathcal{B}=-686\ \text{meV~nm}^2$, 
$\mathcal{C}=0$, 
$\mathcal{D}=-512\ \text{meV~nm}^2$, 
$\mathcal{M}=-10\ \text{meV}$ in the absence of an external magnetic field ($B=0$).  
We find that an HgTe/CdTe QW with $W=200$ nm presents a gap of about $0.44$ meV.
The gap tends to close as we increase the width and reaches values as small as $10^{-5}$ meV 
for $W=1000$ nm.

We address the charge transport properties of the HgTe/CdTe QWs using the Landauer approach. \cite{Datta1997,Haug2008}
In the case of a vanishingly small source-drain bias, the zero temperature conductance of the system 
reads $G=(e^2/h)T(E_F)$, where $T(E_F)=T_{\uparrow}(E_F)+T_{\downarrow}(E_F)$ is the total 
transmission between the left and right contacts at the Fermi energy $E_F$ and
\begin{align}
    T_{\sigma}(E_F) = \text{Tr} \left[ \Gamma_1(E_F) G^r_{\sigma}(E_F) \Gamma_2(E_F) G^a_{\sigma}(E_F) \right]
\end{align}
are the transmissions for each spin $\sigma=\uparrow,\downarrow$.
Here $G^r_{\sigma}$ ($G^a_{\sigma}=\left[G^r_{\sigma}\right]^\dagger$) is the retarded 
(advanced) Green's function for charge carriers with spin $\sigma$ and $\Gamma_1$ 
($\Gamma_2$) is the (spin-independent) line-width function accounting for the injection 
and life-time of the carriers states in the left (right) contact. 

The discrete model Hamiltonian  presented above allows for a very amenable implementation of the recursive Green's functions technique.\cite{Lewenkopf2013}
We compute the line-widths $\Gamma_1$ and $\Gamma_2$ with standard decimation 
methods \cite{LopezSancho1985} and the full retarded Green's function $G^r_{\sigma}$  
in the system central region using the RGF. \cite{Lewenkopf2013} 
We gain additional insight by computing the local transmission 
\begin{align}
    T_{i,j;\sigma}^\alpha(E_F) = -2\text{Im} \left\{ \left[ G^r_{\sigma} \Gamma_\alpha G^a_{\sigma} \right]_{i,j} t_{j,i} \right\}
        \label{localT}
\end{align}
between two neighboring states $i$ and $j$ connected by the hopping matrix 
element $t_{i,j}$ for the charge current injected from the contact $\alpha=1,2$.

\begin{figure}[t]
\begin{center}
\includegraphics[width=1.0\columnwidth]{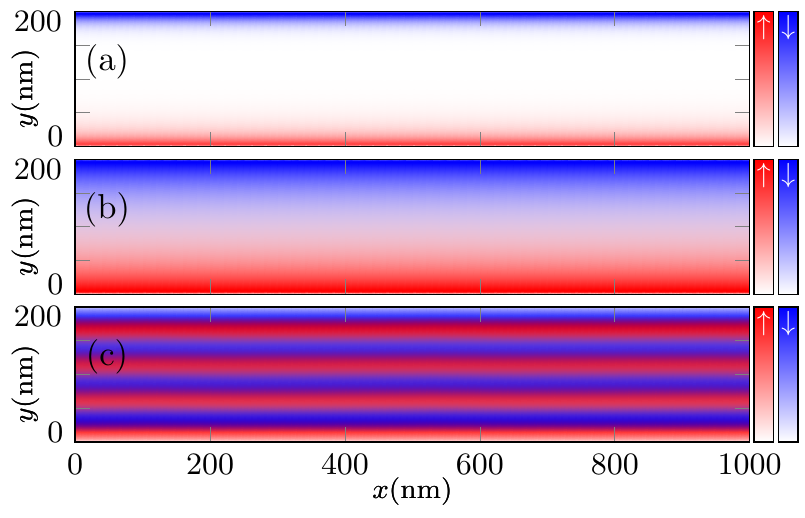}
\caption{ Local currents in a homogeneous ($eV_g\!=\!0$) $200 \times 1000$ nm HgTe/CdTe QW for different values of the Fermi energy $E_F$ in the contacts, namely (a) $E_F=0$ meV, (b) $E_F=10$ meV, (c) $E_F=25$ meV, and $B\!=\!0$. 
Spin-polarized edge transport (from left to right) is evident for $E_F$ inside the gap. 
}
\label{fig:LocCurrentsVgeq0Beq0}
\end{center}
\end{figure}

\section{Results}\label{sec:Results}

In this section, we analyze the magnetotransport properties of homopolar and heteropolar junctions 
in HgTe/CdTe QWs by studying the local transmission of different possible double junction system 
configurations, namely, n-n$^\prime$-n, n-TI-n (n-edge-n for $B\!\neq\!0$), and n-p-n junctions. 
We present separately the analysis of the cases of $B\!=\!0$ (Sec.\ \ref{sec:Beq0results}) and 
$B\!\neq\!0$ (Sec.\ \ref{sec:B05results}).

\subsection{Zero magnetic field}\label{sec:Beq0results}

Here we study the charge transport through n-n$^{\prime}$-n, n-TI-n, and n-p-n junctions in the presence of TRS, that is $B\!=\!0$. 
As mentioned previously, we consider the case of an inverted HgTe/CdTe quantum well, $\mathcal{M}\!<\!0$ in Eq.~\eqref{eq:Hamiltonian}.
Such systems support topologically-protected edge states when TRS is preserved. Thus, the ``edge'' portion of the junction represents a topological insulator.

Let us begin discussing the $eV_g\!=\!0$ case, where the system is uniform. 
Since here the spectrum is known (e.g., Fig.\ \ref{fig:dispersion1D}), the current profile 
serves to test the accuracy of our results and to introduce the tools we use in this study.

Using Eq.~(\ref{localT}) we calculate the stationary local left-to-right transmission between the sites 
$i$ and $j$ as $T(x_{ij},y_{ij}) \equiv \left| T^{\alpha=1}_{i,j;\uparrow}(E_F) + T^{\alpha=1}_{i,j;\downarrow}(E_F) 
\right|$ where ($x_{ij},y_{ij}$) is the midpoint between the sites. 
For each $E_F$ value, we plot a color map 
of the normalized left-to-right transmission $\bar{T}_{\sigma}(x_i,y_j) \equiv \left(T(x_{ij},y_{ij})/T_{\rm max}\right) \eta_\sigma$ where $T_{\rm max}$ is the maximum value of $T(x_{ij},y_{ij})$ and $\eta_\sigma$ is the fraction of $T(x_{ij},y_{ij})$ composed by the spin $\sigma=\uparrow,\downarrow$ component. In this scheme, the values of $\eta_\sigma$ belong to the interval $[0,1]$ satisfying $\eta_\uparrow + \eta_\downarrow = 1$. Thus, $\bar{T}_{\sigma}(x_i,y_j)$ for each $\sigma$ varies between 0 and 1, with the unit representing full spin polarization and maximum transmission.

\begin{figure}[t]
\begin{center}
\includegraphics[width=1.0\columnwidth]{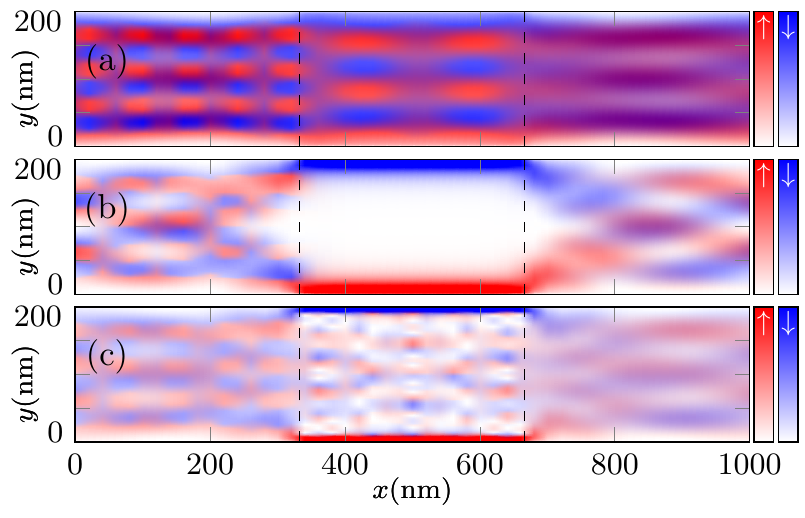}
\caption{
(Color online) 
Local currents in different junctions: (a) n-n$^\prime$-n  (b)  n-TI-n, and (c) n-p-n  junctions. 
In all cases, we considered a 
$200 \times 1000$ nm HgTe/CdTe QW and $B\!=\!0$. The gate voltage in the central region was held, respectively at (a) $eV_g=10$ meV (b) $eV_g=30$ meV and (c) $eV_g=60$ meV. The $E_F$ in the contacts was set to $30$ meV from the bottom of the valence band.  Notice the edge-dominated transport in the n-TI-n junction, for which $E_F$ is set inside the gap in the central region. 
}
\label{fig:LocCurrents3junction}
\end{center}
\end{figure}

Typical results for $\bar{T}_{\sigma}(x_i,y_j)$ are shown in Figure \ref{fig:LocCurrentsVgeq0Beq0}.
For $E_F$ within the gap (Fig.~\ref{fig:LocCurrentsVgeq0Beq0}a), there is only a single pair of states crossing the 
Fermi energy, which are localized at the QW edges. Thus, 
the  current is carried 
by edge states with the expected spin texture of a topological insulator.
As $E_F$ is tuned closer to the bottom of the conduction band, the local currents still flow mostly through spin-polarized states near the edges but the contribution from bulk states become more prominent, as shown in Fig.~\ref{fig:LocCurrentsVgeq0Beq0}b.

In the n and p regions ($E_F$ above and below the gap, respectively) there are well-defined spin-polarized stripes of current through the bulk. This is an interesting pattern: It implies a 
spatial separation of the spin-polarized currents through the bulk. 
This pattern originates from the different pairs of bulk and edge states crossing the Fermi level 
with positive group velocity, $v_k=\partial E(k)/\partial k >0$. The helical nature of the states 
implies that each pair will have opposite spins polarizations. Moreover, the states in each pair 
are mostly symmetrically localized around the center of the strip, in opposite sides of the system, 
creating the pattern shown in Fig.~\ref{fig:LocCurrentsVgeq0Beq0}c.

We now turn to the $eV_g\neq 0$ case. 
Depending on the magnitude of $V_g$, we model a n-n$^\prime$-n junction 
(Fig.~\ref{fig:LocCurrents3junction}a), a n-TI-n junction (Fig.~\ref{fig:LocCurrents3junction}b), or a n-p-n 
heteropolar junction (Fig.~\ref{fig:LocCurrents3junction}c). 
As we discuss below, these junctions are characterized
by  a very distinct current density flow behavior.

\begin{figure}[t]
\begin{center}
\includegraphics[width=0.85\columnwidth]{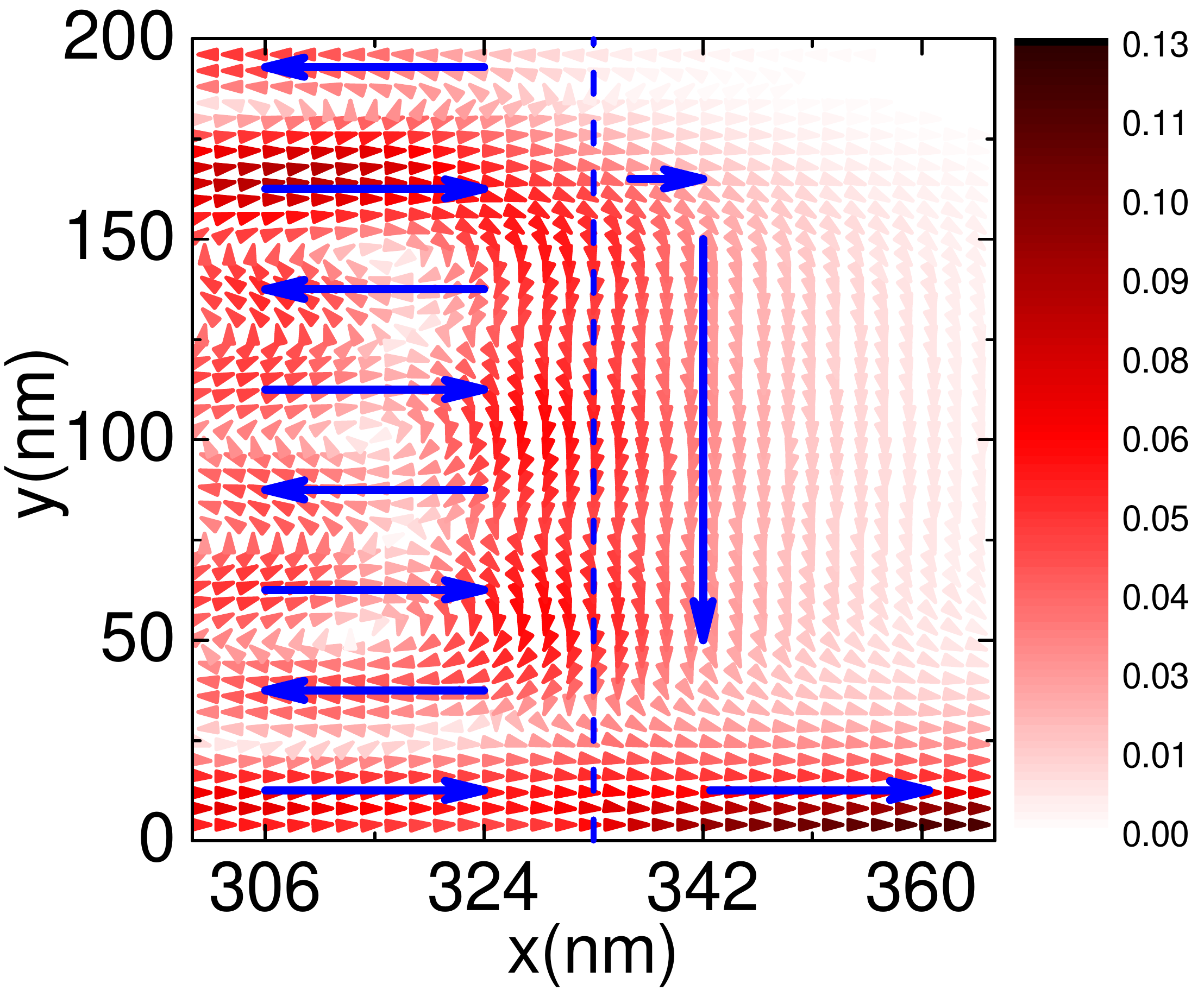}
\caption{ (Color online) Detail of the spin \textcolor{blue}{up} current along the interface 
in an n-TI junction, showing a ``snake-like" pattern in the ``n side'' (left) of the junction. 
In the ``TI side" (right), {the spin up electrons flow parallel to the interface to the bottom edge state, where the left-right transmission across the junction takes place.}
}
\label{fig:LocCurrentsnTInjunction}
\end{center}
\end{figure}

Now we fix $E_F$ at $30$ meV to study n-n$^\prime$-n junctions (Fig.~\ref{fig:LocCurrents3junction}a). 
The current flow shows spin-polarized stripes 
across the QW transverse direction, similar to those observed for $E_F$ outside the gap in the 
$eV_g=0$ case. 
The situation is different in the n-p-n configuration (Fig.~\ref{fig:LocCurrents3junction}c).  
Here, the stripe pattern in the central region seen in the n-n$^{\prime}$-n junction vanishes due to the 
spatial mismatch between n-type and p-type states with positive group velocity. 
As a consequence of this mismatch, in the p-doped region the electronic transport is concentrated at 
the system edges, even though there are bulk states crossing the Fermi energy.

The n-TI-n configuration (Fig.~\ref{fig:LocCurrents3junction}b) shows a ``spatial filtering",
where the current flows through spin-polarized edge states.
Interestingly, reflections at the n-edge interface create a ``snake-like" pattern for the spin-polarized currents. 

This is better illustrated by Fig.~\ref{fig:LocCurrentsnTInjunction}, where the spin up
component of the transmission near the interface is shown for clarity. 
As previously discussed, in the TI region the spin up current is localized at the bottom edge. 
This behavior becomes increasingly clear as one moves away from the interface. 
Fig.~\ref{fig:LocCurrentsnTInjunction} also shows a strong downward flow of spin up electrons 
\textit{parallel to the interface}, represented by the (blue) vertical arrow.
Spin up electron injected in the upper part of the junction cannot propagate the TI region and
move along snake-like trajectories along the interface, \cite{Carmier2010,Carmier2011}
that channels the flow towards the system bottom edge. 

On the n-doped side of the junction the behavior is strikingly different. The spin up electrons 
flow alternates in direction along the system transversal direction. As above, the bottom edge 
states  also contributes  to the 
{left-right} current in the n-region, producing a strong left-to-right spin up component 
matching the flux on the TI side. By contrast, the contribution from bulk states is either 
(i) reflected at the interface, producing small vortex-like patterns and a backward flow, or 
(ii) injected in the TI region in the upper section and then channelled downward along the interface. 
The combination of these two effects produces the current pattern in the n-doped region indicated by the 
(blue) horizontal arrows in Fig.~\ref{fig:LocCurrentsnTInjunction}.

\begin{figure}[t]
\begin{center}
\includegraphics[width=0.85\columnwidth]{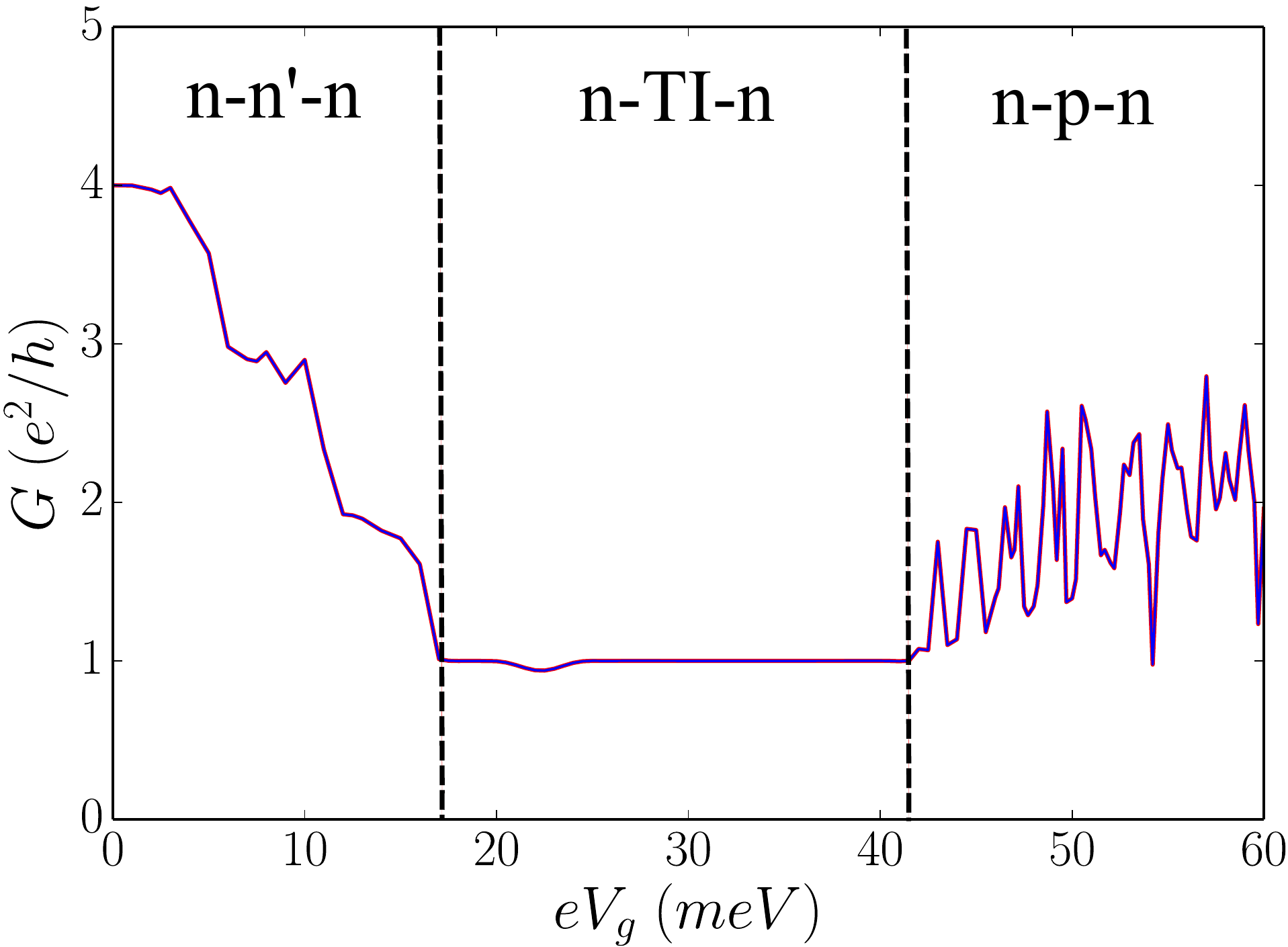}
\caption{ Landauer conductance as a function of $eV_g$ for a $200 \times 1000$ nm 
HgTe/CdTe ribbon for $B\!=\!0$. Transport in the n-n$^{\prime}$-n shows plateaus, with 
small oscillations, while the n-TI-n region is characterized by a clear $2e^2/h$ plateau. 
The little bump at $\approx 22.5$ meV signals a small gap in the spectrum shown in Fig.\ \ref{fig:dispersion1D}.
Strong oscillations in the n-p-n region occur due to the mismatch of the wave-functions in n and p regions.
}
\label{fig:TotCurrentBeq0}
\end{center}
\end{figure}

This picture allows us to interpret the conductance in these systems. 
Figure~\ref{fig:TotCurrentBeq0} shows the conductance per spin as a function of the gate voltage $V_g$. 
The conductance plateaus in the n-n$^{\prime}$-n region essentially count the number open modes 
at the Fermi energy in the central region for a given $V_g$. As $V_g$ is tuned so that the $E_F$ lies 
inside the gap, a clear $2e^2/h$ plateau appears. A small depression in the transmission near 
$eV_g\approx 22.5$ meV signals the presence of a finite-size gap in the spectrum. The gap is small 
enough so that the effective broadening arising from the coupling of the system to the contacts 
(which is captured by the RGF approach) is sufficient to give a large contribution to the transmission 
at that energy value.
In the n-p-n region, the conductance oscillates rapidly with $V_g$. 
This is a result of the multiple reflections and the wave mismatch between n and p regions. Note that  
the states with positive group velocity, that contribute to the left-to-right charge flow, have opposite 
phase velocity in n and p regions which enhance the mismatch between those states.

\begin{figure}[t]
\centering
\includegraphics[width=1.0\columnwidth]{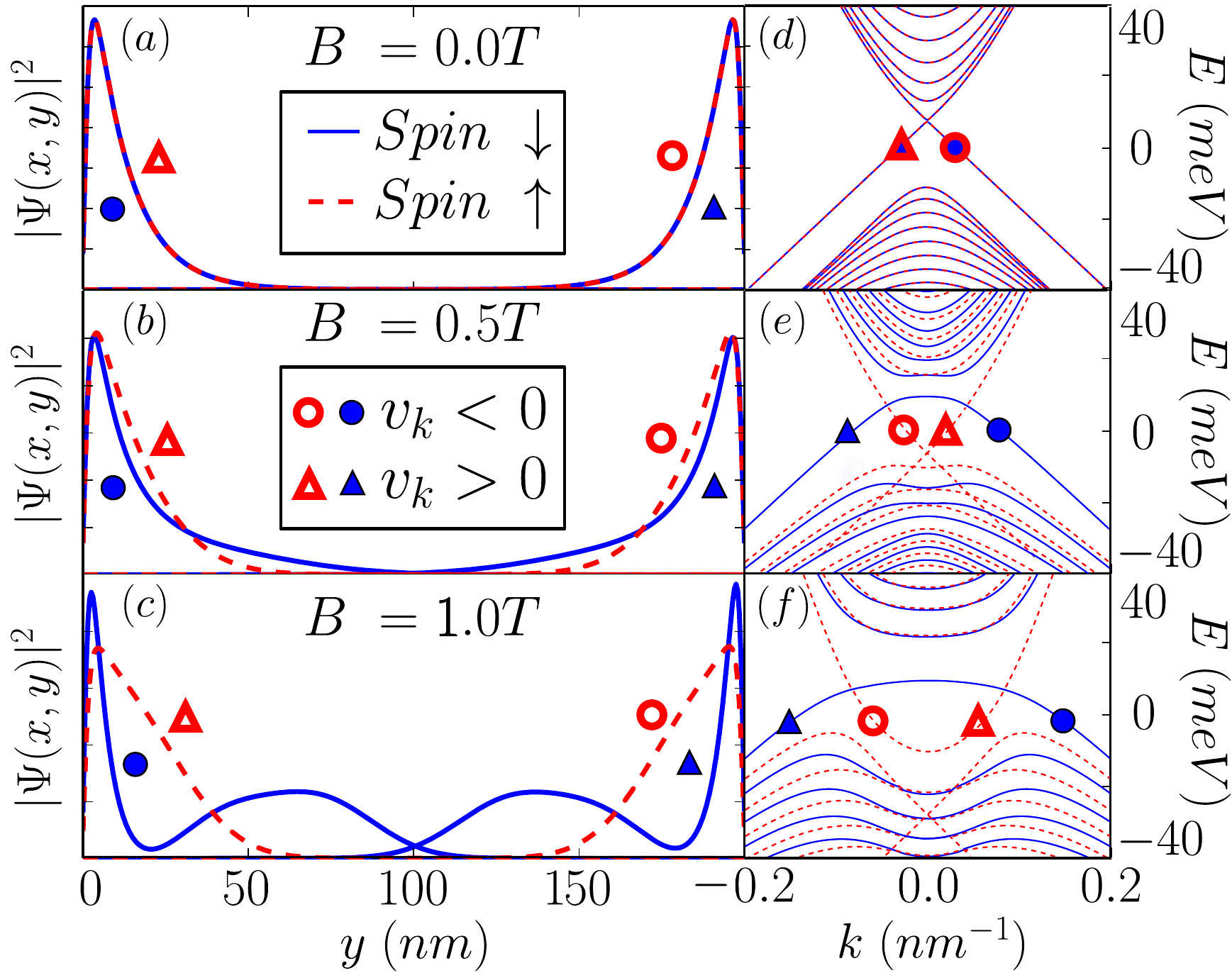}
\caption{(Color online) Wave functions for states inside the bulk gap for $L=200$ nm and 
(a) $B=0$, (b) $0.5$, and (c) $1.0$ T for different $k_x$ values, as marked on the corresponding 
dispersion relations (d) to (f). 
As $B$ increases, the spin up states remain localized close to the edge, while the spin down 
states move toward the bulk. The resulting hybridization opens a gap in the spin down spectrum.
Red dashed (blue solid) lines represent spin up (down) states.
Backward- $(v_k\!<\!0)$ and forward-moving $(v_k\!>\!0)$ states are marked by circles and triangles respectively. 
Filled (empty) symbols represent spin down (up) states.
}
\label{fig:wavefuncBneq0}
\end{figure}

\subsection{Transport at non-zero field }
\label{sec:B05results}

We now study the transport properties of HgTe QW junctions under an external 
perpendicular magnetic field $B$. We consider QWs of $W\!=\! 200$ nm. In this case, the transition to a regime 
where transport is dominated by quantum Hall-like 
chiral edge modes occurs at $B_c \approx 7 - 8$ T.\cite{Scharf:Phys.Rev.B:75418:2012} 
Thus, we restrict our analysis to $B$-fields up to $2$T, where counter-propagating helical 
states are still present in the system. 

\begin{figure}[t]
\begin{center}
\includegraphics[width=1.0\columnwidth]{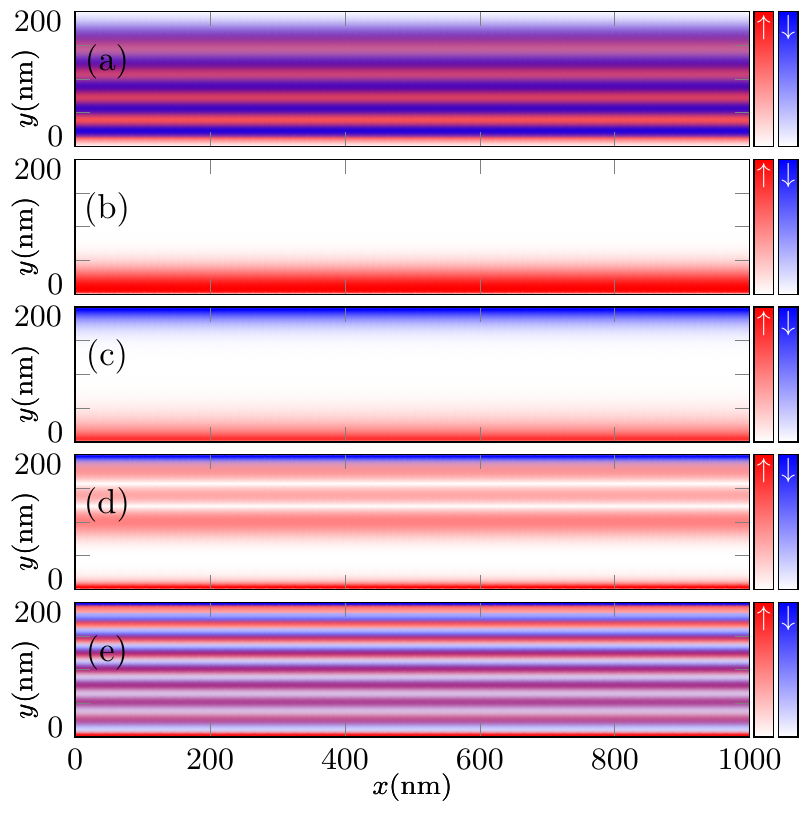}
\caption{ Local currents for an homogeneous ($eV_g=0$) $200 \times 1000$ nm HgTe/CdTe QW subjected 
to a perpendicular magnetic field at $B\!=\!0.5$T. The currents are shown for different values of the Fermi 
energy $E_F$ in the contacts, namely (a) 40 meV (n-type transport), (b) 15 meV, (c) 5 meV, (d) -16 meV, 
and (e) -40 meV (p-type transport) (see Fig.\ \ref{fig:dispersion1D}). 
}
\label{fig:SpectrumLocCurrentsBeq05}
\end{center}
\end{figure}

As it is well known,\cite{Hasan:3045:2010,Qi:1057:2011,Ando:J.Phys.Soc.Jpn.:102001:2013} a magnetic field breaks TRS and thus the edge states lose their topological protection. 
Moreover, the Zeeman term in Eq.~\eqref{eq:Hamiltonian}, although small, also breaks the spin degeneracy. 
The combination of these two effects substantially changes the spin-dependent transport properties across the junction.

We begin by exploring the non-zero $B$ case for the $eV_g\!=\!0$ case.
Figure~\ref{fig:wavefuncBneq0} contrasts the probability distributions of the system states at $E_F=0$ and 
the dispersion relations for representative values of $B$. 
Consistent with previous studies, \cite{Chen:Phys.Rev.B:125401:2012} for $B=0.5$T, 
a  well-pronounced gap ($\sim 10$ meV) appears for spin down states, while the spin up
 states show no-gap.
As the field increases, the  probability density of the spin up states remains concentrated at the edges, while the spin down states penetrate deeper into the bulk.
This behavior is consistent with the local currents shown in  Figs.~\ref{fig:SpectrumLocCurrentsBeq05}b 
to \ref{fig:SpectrumLocCurrentsBeq05}d for selected values of the Fermi energy. In those cases, the 
asymmetry with respect to the $y$ axis (across the width) appears because only the forward moving states at one edge are present in the transport.

For n- and p-type transport Figs.~\ref{fig:SpectrumLocCurrentsBeq05}a and
 \ref{fig:SpectrumLocCurrentsBeq05}e, respectively, the bulk currents flow along
nearly spin-polarized stripes, similarly to the $B=0$ case. 
However, some interesting differences appear. When the Fermi energy lies inside 
the spin down gap, the current is fully spin-up polarized, flowing through the lower 
edge (Fig.~\ref{fig:SpectrumLocCurrentsBeq05}b). 
As $E_F$ is tuned slightly below the spin up gap, the system shows 
spin-polarized transport on \textit{both} edges, similar to the topological case,
as shown in Fig.~\ref{fig:SpectrumLocCurrentsBeq05}c. 
Note that the threshold for spin up bulk states is higher in 
energy than the spin down states, leading to a region where we have transport 
dominated by \textit{bulk spin up} and \textit{edge spin down} currents (Fig.~\ref{fig:SpectrumLocCurrentsBeq05}d).  

\begin{figure}[t]
\begin{center}
\includegraphics[width=1.0\columnwidth]{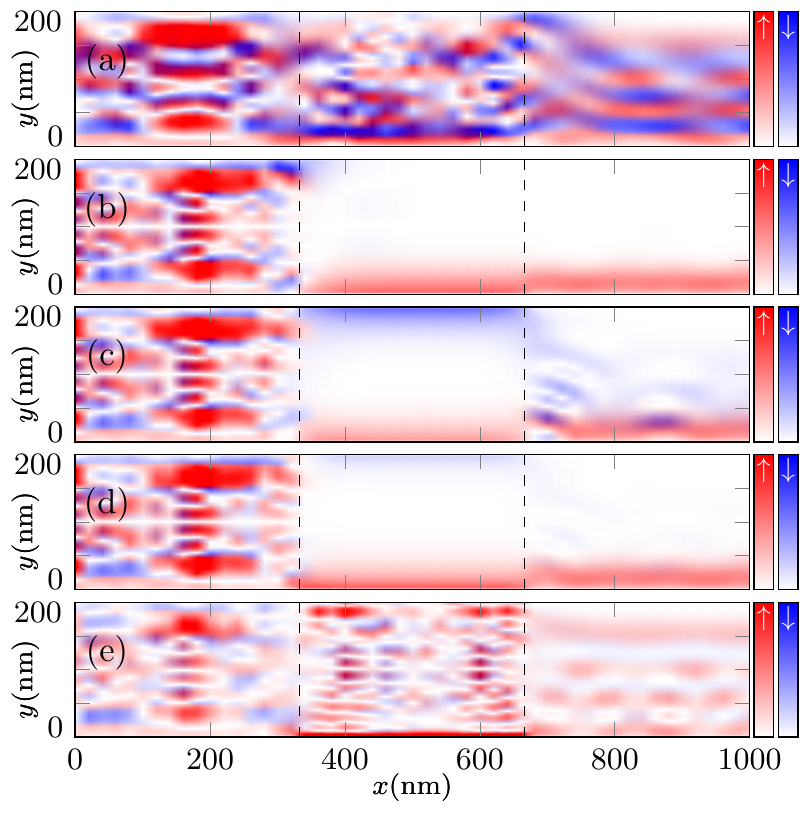}
\caption{Local currents for a n-(central region)-n junction in a  $200 \times 1000$ nm HgTe/CdTe QW, with $B=0.5$T 
and $E_F=30$ meV for different  $eV_g$ values, namely (a) -10 meV (n-n$^\prime$-n junction), (b) 20 meV, (c) 22.5 meV,
(d) 25 meV, and (e) 60 meV (n-p-n junction).
}
\label{fig:B05CurrentOscillations}
\end{center}
\end{figure}

We now consider the local currents in HgTe n-(central)-n junctions at a finite magnetic field. 
Figure \ref{fig:B05CurrentOscillations} shows the behavior for different values of $V_g$ such 
that $E_F$ lies close to the spin down local gap in the central region. 
Figures \ref{fig:B05CurrentOscillations}a and \ref{fig:B05CurrentOscillations}e correspond 
to n-n$^\prime$-n and n-p-n junctions, respectively. In these cases the transport properties 
are dominated by bulk states and orbital interference effects.

When $E_F$ lies within the spin down gap, a spin up polarized current flows through the 
lower edge of the central region (Fig.~\ref{fig:B05CurrentOscillations}b) and it is injected in 
the right n region through an edge state. A slight increase in $eV_g$ (from 20 to 22.5 meV) 
is sufficient to bring $E_F$ to cross the first spin down edge state below the gap, thereby 
allowing spin down transport through the upper edge of the central region 
(Fig.~\ref{fig:B05CurrentOscillations}c).

Surprisingly, a further small increase in $eV_g$ (from 22.5 to 25 meV) causes the spin 
down current in the central region to practically \textit{vanish}, as shown in 
Fig.~\ref{fig:B05CurrentOscillations}d. This is at odds with the homogeneous case 
(Fig.~\ref{fig:SpectrumLocCurrentsBeq05}) where spin down currents are always 
present as long as $E_F$ lies outside the spin down gap.
We attribute this suppression to the large change in momentum across the n-edge 
junction necessary for the propagation of spin down electrons in the central region, as
inferred from the band structure in Fig.~\ref{fig:wavefuncBneq0}e.

\begin{figure}[t]
\begin{center}
\includegraphics[width=\columnwidth]{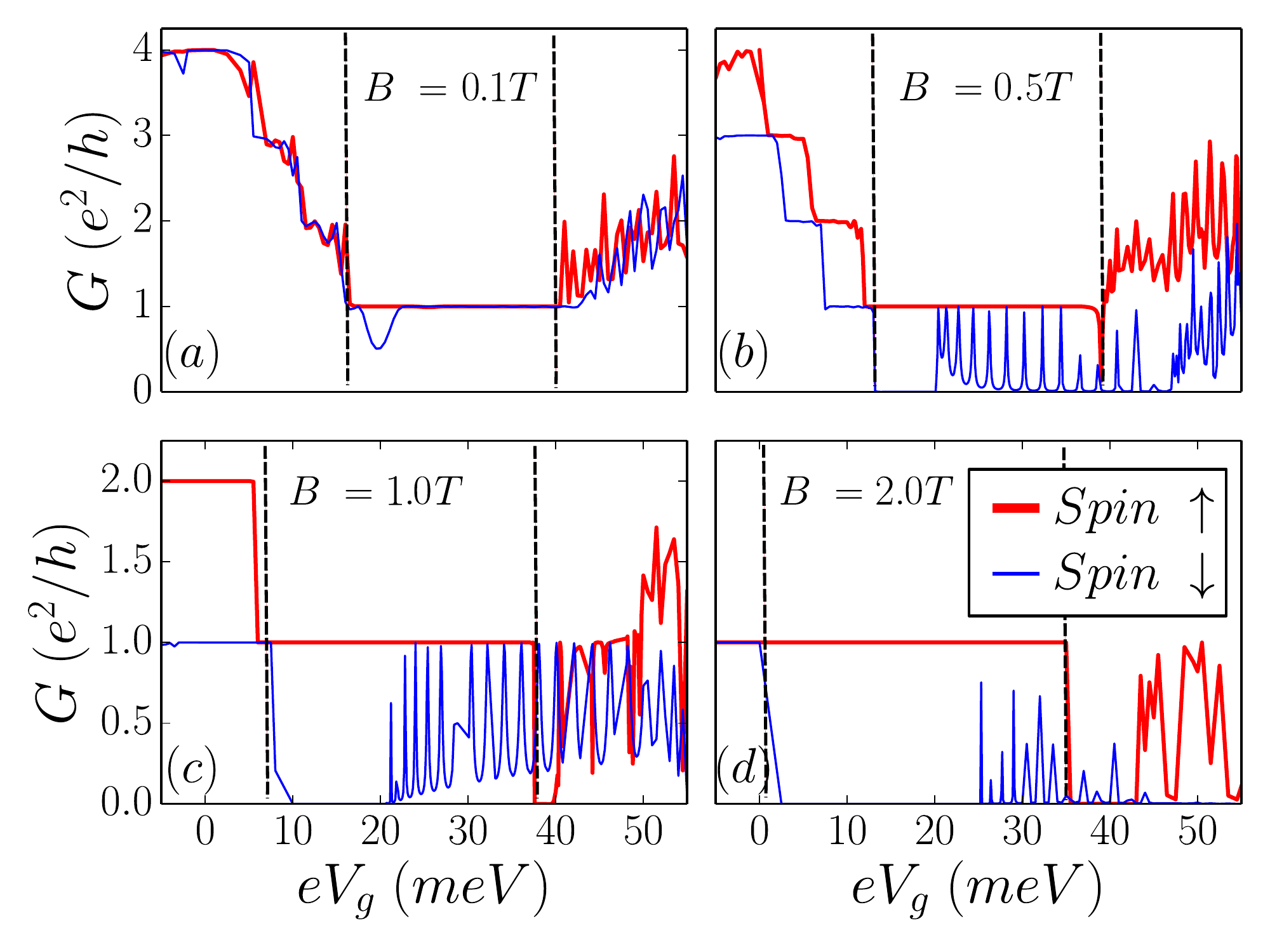}
\caption{ Conductance per spin versus $V_g$ 
for different values of the magnetic field.  The $eV_g$ range sweeps the junction from n-n$^\prime$-n  ($eV_g=10$ meV) to n-p-n ($eV_g=60$ meV). For weak fields, the currents are nearly spin independent,  with a $2e^2/h$ plateau 
due to edge state transport. A gap in the spin down spectrum becomes more prominent for larger fields. For $B\!=\!0.5$T, a clear oscillatory pattern (on/off) in the spin down current appears between the spin gap edge ($eV_g=20$ meV) and the onset of n-p-n behavior ($eV_g \approx 40$ meV). Notice that this edge current behavior extends to fields as large as $B=2$T, panel (d). 
}
\label{fig:CurrentOscillationsII}
\end{center}
\end{figure}

Let us now examine the conductance across the junction as a function of the gate voltage $V_g$ for $B=0.5$ T.
Figure \ref{fig:CurrentOscillationsII} shows a clear oscillatory pattern of the spin down current 
for $eV_g>20$ meV up to the onset of n-p-n behavior at $eV_g \approx 40$ meV. In the same 
$V_g$ range, the spin up conductance remains at a $e^2/h$ plateau, indicating  spin-polarized 
edge transmission for the $V_g$ values where the spin down current essentially vanishes.

\begin{figure}[t]
\begin{center}
\includegraphics[width=0.8\columnwidth]{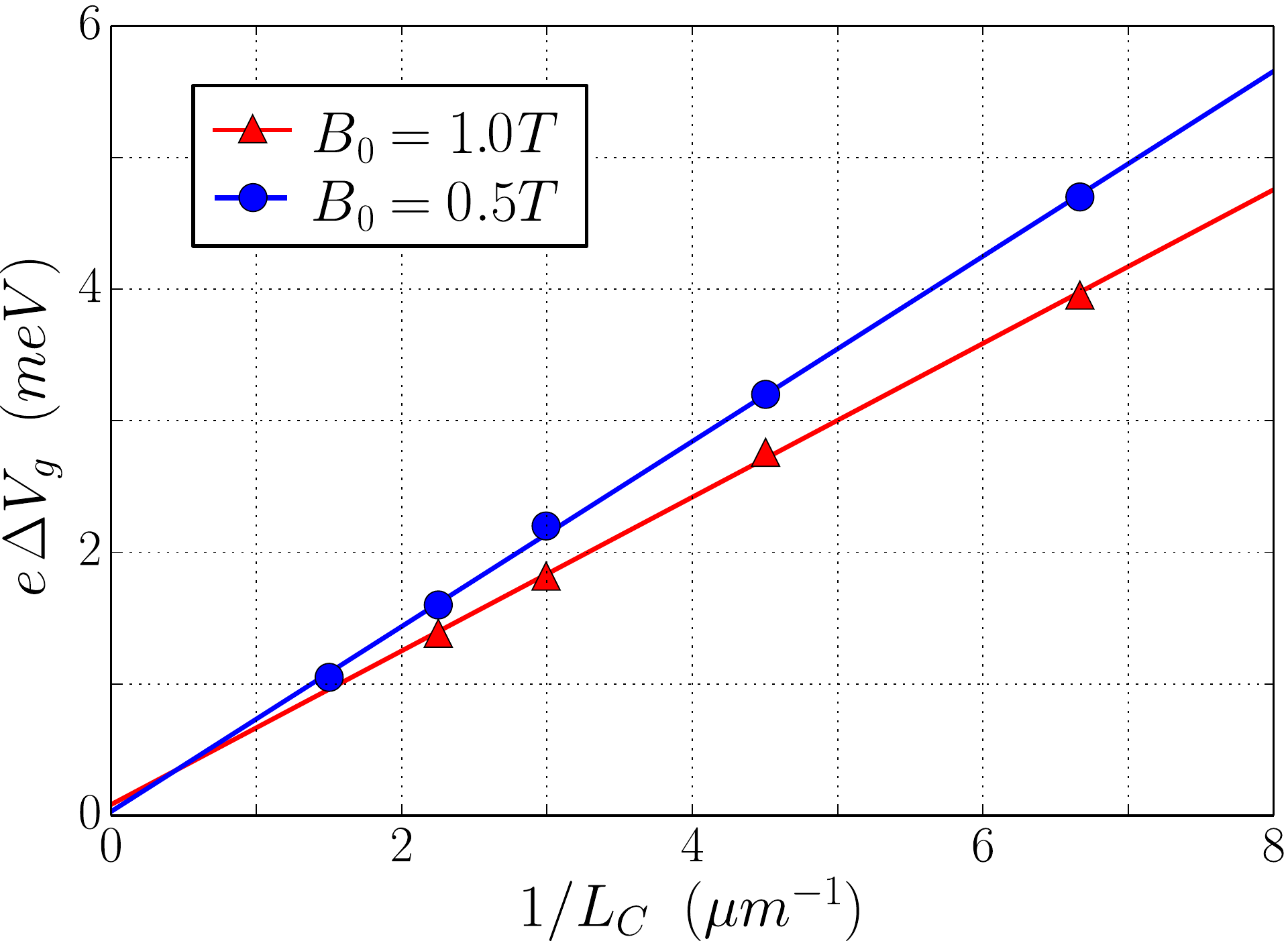}
\caption{ Spin-down transmission peak spacing versus $1/L_C$ ($L_C$ is the central region length) for different magnetic field values. The $1/L_C$ scaling in all cases is consistent with Fabry-P\'erot-like oscillations due to backscattering via reflection with the step potential at the junction interface.
}
\label{fig:FabryPerot}
\end{center}
\end{figure}

We associate these peaks with Fabry-P\'erot resonances caused by backscattering at the junction 
interfaces. Figure~\ref{fig:FabryPerot} shows that the spacing $\Delta V_g$ between the transmission 
peaks displays a linear scaling with the inverse of the central region length $L_C$, indicating single-particle 
interference due to backscattering at the interfaces. Similar phenomena has been investigated 
previously in two-terminal devices in the presence of a magnetic field. 
\cite{Tkachov:Phys.Rev.Lett.:104:166803:2010,Soori:Phys.Rev.B:86:125312:2012}
Here, the presence of the interfaces magnifies the effect, leading to strong oscillations where a perfect spin-polarized transport across the junction is possible.

Fabry-P\'erot-like oscillations also appear for larger magnetic field values and are suppressed for lower ones.
In fact, the oscillations seem to occur only in the presence of a fully developed gap in the spin down spectrum, as shown in Figs.~\ref{fig:CurrentOscillationsII}b and \ref{fig:CurrentOscillationsII}c. 

For $B=2$T, Fig.~\ref{fig:CurrentOscillationsII}d, the behavior is similar. Since the spin down gap is significantly larger, 
the range of $V_g$ 
for which the current displays full spin up polarization  increases from $eV_g \approx 12.5-20$ meV for $B\!=\!0.5$T to $eV_g \approx 5-25$ meV for $B\!=\!2$T. Interestingly, for larger $V_g$ values ($eV_g \gtrsim 35$meV), spin up drops to zero and full spin down polarization is now possible. Thus, for these moderate magnetic field values, the junction operates as a gate-tunable spin polarization switch.

\section{Concluding remarks}\label{sec:Conclusions}

In this paper we have theoretically studied the spin-dependent local currents in HgTe/CdTe quantum-well 
monopolar and heteropolar junctions.  We considered the dependence of the transport properties with an applied magnetic field 
perpendicular to the sample and the resulting transition from topologically protected edge transport 
to a regime where spin backscattering is allowed at the junction barriers.

For zero magnetic field, our recursive Green's functions calculations for the local currents show distinct characteristics for n-n$^\prime$-n, n-TI-n and n-p-n junctions. While the bulk contributions to the current are strong in monopolar (n-n$^\prime$-n) junctions, in the heteropolar case transport is dominated by edge states in the central region. In n-TI-n junctions, the spin-resolved flow alternates in direction along the system transverse direction in the n-side of the first n-TI interface. By contrast, the TI side shows  currents flowing \textit{parallel} to the interface toward the edges, where the main flow occurs.

Edge states still give a strong contribution to the transmission in the presence of TRS-breaking perpendicular magnetic field. Interestingly, the magnetic field opens a gap for one of the spins. Quantum interference due to backscattering at the interface produces spin-resolved Fabry-P\'erot-like oscillations in the transmission as a function of the gate applied to the central region.  

The combination of the gap opening and the Fabry-P\'erot oscillations for only one of the spins allows for the production of tunable spin-polarized currents across the junction for moderate ($B<1$T) values of the magnetic field. We stress that these results are generic for other inverted QWs displaying 2D topological insulator behavior such as InAs/GaSb.\cite{Du:Phys.Rev.Lett.:096802:2015,Karalic:Phys.Rev.Lett.:206801:2017} This opens the prospect for applications of inverted QW heteropolar junctions in spintronic devices.

\acknowledgments 
DN\ and LRFL\ acknowledge support from the Brazilian funding agencies CNPq, FAPERJ and Capes.
LDS\ acknowledges support from CNPq grants 307107/2013-2 and 449148/2014-9, and FAPESP grant 2016/18495-4. 
CHL is supported by CNPq grant 308801/2015-6 and FAPERJ grant E-26/202.917/2015. 


%

\end{document}